\documentclass[10pt,twocolumn]{article}

\usepackage[margin=0.75in,columnsep=0.25in]{geometry}
\usepackage[T1]{fontenc}
\usepackage[utf8]{inputenc}
\usepackage{microtype}

\usepackage{amsmath}
\usepackage{amssymb}
\usepackage{dsfont}
\usepackage{graphicx}
\usepackage{tikz}

\usepackage{booktabs}
\usepackage{multirow}
\usepackage{array}
\usepackage{makecell}
\usepackage{tabularx}
\usepackage{adjustbox}
\usepackage{etoolbox}

\usepackage{subcaption}

\usepackage[hidelinks]{hyperref}

\AtBeginEnvironment{table}{\small\centering}
\AtBeginEnvironment{table*}{\small\centering}

% ...
% ...

\title{\Large\bfseries
MeMark: Membrane-Space Watermarking for Spiking Neural Networks
}

\author{%
\small
\textbf{Roberto Riaño}$^{1,2}$ \quad
\textbf{Gorka Abad}$^{3}$ \quad
\textbf{Stjepan Picek}$^{1,4}$ \quad
\textbf{Aitor Urbieta}$^{2}$
\\[0.5em]
\footnotesize
$^{1}$Radboud University, The Netherlands
\quad
$^{2}$IKERLAN Technology Research Centre, Spain
\\\footnotesize
$^{3}$University of Bergen, Norway
\quad
$^{4}$University of Zagreb, Croatia
\\[0.4em]
\scriptsize
\texttt{roberto.rianohidalgo@ru.nl} \quad
\texttt{gorka.abad@uib.no}\quad
\texttt{stjepan.picek@ru.nl} \quad
\texttt{aurbieta@ikerlan.es}
}

\date{}

\begin{document}

\maketitle

\begin{abstract}
Spiking Neural Networks (SNNs) are increasingly distributed as pretrained checkpoints and reused as backbones for new tasks. However, current SNN watermarks are mainly verified against the model output. Thus, a user who replaces the output head can keep most of the original network while removing the evidence used for verification. We present MeMark, a watermark designed for the checkpoint-reuse setting. Instead of storing the watermark in the output head, MeMark embeds a multi-bit identifier in the internal membrane state of selected Leaky Integrate-and-Fire (LIF) neurons. A secret input drives each selected neuron to the chosen side of its own firing threshold, and the same threshold is later used to recover the secret bit, so the verifier does not need a learned decoder. 
We evaluate MeMark across recurrent, convolutional, residual, and transformer SNNs. On a 215.4M-parameter SpikeGPT checkpoint, all 20 independent 64-bit keys pass the fixed 51/64 verification rule, while none of the $30\,000$ fresh random keys pass when tested against all 20 protected checkpoints and the clean model. All 20 genuine keys also remain above the threshold after fine-tuning, 90\% pruning, int8 quantization, and output-head replacement. Under our stated threat model, adaptive attacks can weaken the watermark but do not remove the ownership evidence in the settings we test. Additionally, we study false ownership claims, key-aware and key-agnostic removal, partial key disclosure, rollback, and extraction into a student. The results show that MeMark can provide evidence of checkpoint derivatives, while being resistant to the adversary's attacks and complete head replacement. All our code and results are available at our repository.
\end{abstract}

\section{Introduction}
\label{sec:intro}

Deep Neural Networks (DNNs) have achieved excellent performance in many machine learning tasks, including computer vision, speech recognition, and natural language processing~\cite{krizhevsky2012imagenet,graves2013speech,devlin2019bert}. However, training high-performance DNNs is computationally expensive and requires large datasets, specialized hardware, and long optimization times, resulting in substantial investment and energy consumption. This has motivated researchers to explore alternative models that can provide similar capabilities with lower energy consumption.

Spiking Neural Networks (SNNs) are one of the main approaches that are being explored~\cite{tavanaei2019deep,eshraghian2023training}. Unlike conventional DNNs, SNNs communicate through spikes over time and remain inactive when they do not produce an output. Their event-driven behavior reduces unnecessary computation and makes SNNs attractive for low-power systems, achieving up to $12.2\times$ better compute energy efficiency than comparable DNNs~\cite{kundu2021spike}. Additionally, in recent years, SNNs have improved considerably in model size and performance. Spiking-YOLO showed that SNNs can be used for energy-efficient object detection~\cite{kim2020spikingyolo}, while residual architectures made it possible to train deeper spiking models~\cite{sewresnet}. More recent work introduced transformer-based SNNs such as Spikeformer~\cite{spikformer}, QKFormer~\cite{qkformer}, and SpikeGPT~\cite{spikegpt}, demonstrating that models can be scaled to 216M-parameter generative language models. Still, training such models, like their traditional counterparts, can require considerable data, computation, and engineering effort.

This makes the trained model an important asset to protect. Once a checkpoint is released, another user can copy the weights, fine-tune the model, replace part of the network, or reuse the backbone of the model for another task. Model watermarking provides a way to protect the models by embedding information that can be later verified in a suspect model. In traditional DNNs, watermarks have been embedded in their weights, internal activations, learned representations, and trigger-dependent outputs~\cite{uchida, deepsigns, diction, sslwm, eaaw}.

However, only a few works have considered watermarking for SNNs. Poursiami et al.~\cite{poursiami} proposed ownership protection for neuromorphic models to require authorization for their use, while SpikeTimer~\cite{spiketimer} embeds a temporal response that can be verified from the model's output. Output verification is useful when the owner has black-box access to the suspect model. However, it is not suitable when an SNN is reused as a backbone. Indeed, replacing the final task-specific layers can completely remove the output watermark while leaving most of the original model unchanged.

In this work, we introduce MeMark, a multi-bit watermark designed to remain inside the spiking computation even when the output head is replaced. MeMark uses the membrane potentials of the spiking neurons to embed the watermark. We assign each watermark bit to a secret layer, neuron, and timestep, making use of their internal firing thresholds as the decoding method. Then, we train the corresponding membrane potential to lie on the required side of that neuron's threshold when the secret input is present. During verification, the owner of the model can present the same secret input and read the selected states, checking each bit of the response.

We use the membrane information in two different ways. The first is a hard read that checks only which side of the threshold the neuron reaches at that timestep, and since the threshold is already part of the neuron, it does not require a learned decoder. The second keeps the distance from the membrane potential to the threshold and uses it as a confidence score. This distance is also useful during the embedding phase, as it provides a training signal before the spike changes. Additionally, MeMark can train an optional output response for cases where the verifier cannot inspect the internal states. We keep this output carrier as an alternative verification interface, but it is expected to disappear if the output head is completely replaced. Unlike output-based SNN watermarks, the primary MeMark response remains in the backbone even after the output head is replaced. Compared to our matched DICTION~\cite{diction} adaptation, MeMark obtains the same on-challenge recovery and comparable fidelity, but does not store a learned projection and reveals less payload information when the challenge is absent.

However, obtaining the correct watermark response is not enough to prove ownership. A claimant with white-box access can inspect a model and craft a new key by choosing coordinates and bit values that already match its internal states. Similar ambiguity and false-ownership attacks have been identified for DNN watermarking~\cite{passport,mor}. We evaluate this case and show that a fitted key can pass the model-level rule. Therefore, before releasing the protected checkpoint, the owner must also publish a timestamped commitment that links the key, owner, release, and the specific checkpoint hash. The watermark shows that the suspect model contains the registered response, while the commitment demonstrates that the key was fixed before the claim.

We evaluate MeMark on recurrent, convolutional, residual, and transformer SNN architectures. Our largest experiment embeds 20 independent 64-bit keys in a 215.4M-parameter SpikeGPT checkpoint, and evaluates the verification rule using $30\,000$ random keys. We study the hard membrane read, membrane-distance score, and the optional output carrier under head replacement, fine-tuning, pruning, quantization, SNN-specific modifications, structural changes, adaptive removal, rollback, partial key disclosure, and extraction. For attacks that modify the checkpoint, we also report the effect on normal task performance.

Our main contributions are as follows:
\begin{itemize}
    \item We introduce MeMark, a challenge-response watermark that embeds a multi-bit identifier in secret LIF states and supports both hard threshold and continuous membrane-distance verification without a learned internal decoder.
    \item We include an optional output carrier for query-only verification and a timestamped release commitment that prevents a key fitted after model inspection from being considered as prior ownership evidence.
    \item We evaluate MeMark across eight SNN testbeds, including 20 independent embeddings in a 215.4M-parameter SpikeGPT checkpoint and $30\,000$ random keys, under checkpoint reuse and several adaptive and model-modification attacks. We also compare to previous watermarking methods, adapt DNN watermarking techniques, and re-implement Poursiami et al.~\cite{poursiami} as an SNN-specific baseline. The comparison shows that MeMark keeps the internal evidence when the output-based SNN watermarks are removed, while avoiding the learned projection used by prior DNN activation-based watermarks.
\end{itemize}

\section{Background}
\label{sec:background}

\subsection{Spiking Neural Networks}
Unlike conventional DNNs, which use continuous-valued activations, SNNs work by aggregating discrete events, or spikes, over time. The output of a spiking neuron depends on both the current input it receives and the information accumulated from previous time steps.

One of the most commonly used models is the Leaky Integrate-and-Fire (LIF) neuron~\cite{hunsberger2015lif}. A LIF neuron keeps an internal value called membrane potential; all incoming spikes increase this value, while the information from previous time steps gradually lowers. When enough spikes gather and the membrane potential reaches its firing threshold $v_{th}$, the neuron produces a spike and its membrane potential is reset. The firing decision can be expressed as

\begin{equation}
    s_i(t)=
    \begin{cases}
        1, & \text{if } v_i(t) \geq v_{th},\\
        0, & \text{otherwise,}
    \end{cases}
    \label{eq:fire}
\end{equation}
where $v_i(t)$ is the membrane potential of the LIF neuron $i$ at timestep $t$. MeMark uses this internal value immediately before the spike reset.

One of the main challenges when training SNNs lies in Eq.~\eqref{eq:fire}: since the spike operation is not differentiable, standard backpropagation cannot be applied. A common solution is surrogate-gradient training~\cite{neftci2019surrogate}, where a differentiable approximation of the binary spikes is used during backpropagation, allowing SNNs to be trained with standard gradient-based optimizers.

\subsection{Model Watermarking}

The objective of model watermarking is to protect the property of a trained model by embedding information that can later be verified in a suspect model. The owner first embeds a secret response while preserving the original task performance. Then, during verification, the same secret information is used to check whether the response is present in the suspect model. 

Watermarks are commonly distinguished by the access required during verification. Black-box methods only require access to the model's inputs and outputs, usually by using a secret trigger that produces a predefined output. Conversely, white-box methods assume access to the model itself and can store information in the weights, normalization parameters~\cite{uchida,passport,passportnorm}, or in internal activations produced by a secret input~\cite{deepsigns,diction}. While black-box verification methods are convenient for remote models or models accessed via API, white-box verification can place evidence deeper inside the network itself.

To our knowledge, only two works have considered watermarking specifically for SNNs, both being verified through the model's output. This is useful when only query access is available, but it becomes unsuitable when an SNN is reused as a backbone and its output layers are replaced. For this reason, MeMark primarily verifies the watermark from internal LIF states (we also keep an optional output response for cases where internal access is not available).
\section{Threat Model}
\label{sec:threat}

\subsection{Deployment Setting}
We consider a scenario where a model owner trains a proprietary SNN and releases it as a pretrained checkpoint that can later be reused or fine-tuned by other users. Before releasing the model, the owner embeds the watermark and keeps the watermark key private. The attacker is a recipient who obtains the protected checkpoints and wants to steal the model to reuse or redistribute it while removing the evidence that connects it to the original release. Once the checkpoint is released, we assume that the owner has no control over how it is used or modified.

The attacker has full access to the released model, including its architecture and weights. Therefore, they can fine-tune, prune, or quantize the model, replace the output head, modify the LIF parameters, reorganize hidden channels, or reuse just a portion of the network. We also assume the attacker knows MeMark has been applied and understands how it works. The only knowledge the attackers lack is the secret key, including the challenge input, selected LIF coordinates, payload bits, and optional output targets. In Section~\ref{sec:adaptive}, we consider stronger attackers that try to search for this information or know part of the secret key.

Watermark verification for MeMark requires white-box access to the suspect model. In an ownership dispute, the verifier must be able to run the secret challenge through the suspected checkpoint and inspect the selected LIF states at the corresponding time steps. If only the model outputs are available, the optional output carrier can still be verified. However, this method is weaker, as replacing the output head can remove it. This is why MeMark embeds the primary watermark deeper, inside the SNN's membrane states.

The owner keeps two pieces of information secret before a dispute. The first one is the secret key, as an attacker who knows the selected coordinates and target bits can directly target those states for removal. The second one is the exact checkpoint before the watermark was embedded. Having both that pre-watermark checkpoint and the protected one would allow the attacker to compare them and search for the changes introduced during embedding. Our robustness claims therefore assume that this exact pre-watermark checkpoint is not available to the attacker. Section~\ref{sec:rollbackresults} evaluates the stronger case where this assumption is broken.

We also consider a different type of attacker who tries to make a false ownership claim over a published model. With white-box access to a model, a dishonest claimant can inspect its internal states and choose a key that already matches them. Section~\ref{sec:falseclaim} shows that this can produce a valid model-level response even though no watermark was embedded by the attacker. For this reason, before releasing the protected checkpoint, the owner must publish a timestamped commitment that binds the watermark key to that specific checkpoint. During a dispute, the claimant must show that the earlier committed key also verifies for the suspect model. This commitment procedure is described in Section~\ref{sec:method}.

Finally, MeMark is intended to identify checkpoint lineage. This includes models that reuse the protected internal computation, even after the checkpoint has been modified. A fresh student trained only from the model's normal outputs is outside this setting, as it may reproduce the task without inheriting the original model's internal response. We explore this case in Appendix~\ref{app:extraction}.
\section{Method}
\label{sec:method}

\begin{figure}[t]
\centering
\includegraphics[width=0.9\linewidth]{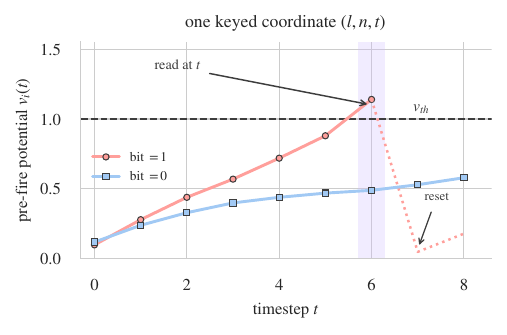}
\caption{MeMark stores each watermark bit in the pre-fire state of a selected LIF neuron. A one-bit is moved above the neuron's firing threshold and a zero-bit below it at the selected time step. The same threshold is used again during verification.}
\label{fig:mechanism}
\end{figure}

This section describes how MeMark embeds and later verifies a watermark into the internal states of an SNN. As discussed in Section~\ref{sec:background}, a LIF neuron keeps a membrane potential and produces a spike when its value reaches the firing threshold. We use this behavior to store the watermark by selecting a small number of LIF neurons and time steps, associating one watermark bit with each selected state. When the secret input is presented, the verifier reads the LIF neurons' membrane potential at the specific time steps and reconstructs the multi-bit payload. A one-bit is represented by a membrane potential above the firing threshold, while a zero-bit by a membrane potential below it, as illustrated in Figure~\ref{fig:mechanism}.

\subsection{Watermark Key}

As shown by Liu et al.~\cite{mor}, a watermark response alone cannot establish when the corresponding ownership evidence was created. They address post-hoc ownership claims using a timestamped cryptographic commitment. We follow a similar approach for MeMark.

Before embedding the watermark, the owner generates a secret input $x_{wm}$, a binary payload $W\in\{0,1\}^{B}$, and the LIF states $C$ that will contain the payload. We denote the complete watermark key as 
\begin{equation}
K=(x_{wm},C,W),
\label{eq:key}
\end{equation}
where
\begin{equation}
C=\{(l_k,n_k,t_k)\}_{k=1}^{B}.
\label{eq:coordinates}
\end{equation}

Each watermark key is generated independently before embedding. The payload bits are sampled uniformly from $\{0,1\}$, and the $B$ carrier coordinates are sampled uniformly without replacement from the eligible pre-fire LIF states. The architecture-specific challenge generation and eligible state sets are described in Appendix~\ref{app:keygen}. Additionally, the effect of selecting nearby states is measured in Appendix~\ref{app:ablations}.

The input $x_{wm}$ determines when the watermark response appears, while $C$ determines where the verifier has to look. Lastly, the payload $W$ determines the expected side of the firing threshold. Thus, observing the model on regular input does not directly reveal the watermark, and knowing the challenge alone is not enough to locate the multi-bit payload.

Once the watermark has been embedded, the owner binds the key to the protected model before releasing it. This is needed for the false-claim scenario described in Section~\ref{sec:threat}. 

Let $id_o$ identify the owner, $id_r$ the release identifier, and $h_M=H(M_K)$ the SHA-256 hash of the protected checkpoint. To avoid ambiguous concatenations, we encode each relevant field together with its length and add a fixed tag before hashing. The owner creates
\begin{equation}
c=H(\texttt{MeMark-Commit1}\Vert\operatorname{Enc}(K)\Vert\operatorname{Enc}(id_o)\Vert\operatorname{Enc}(id_r)\Vert h_M).
\label{eq:commitment}
\end{equation}
The key is generated with at least 128 bits of entropy using a cryptographically secure random number generator.

The owner then submits $c$ to an external timestamp service before publishing the checkpoint. During an ownership claim, the owner reveals these values, and the verifier checks that the commitment was created before the suspect model was released, ensuring that the key was not crafted afterward to fit the model. 

\subsection{Membrane Watermark Embedding}
To embed each bit, we modify the membrane potential of its selected neuron when the secret input is presented. Considering a bit $W_k$ and its coordinate $(l_k,n_k,t_k)$, we denote the pre-fire membrane potential at this position as
\[
v_k=v_{l_k,n_k}(t_k),
\]
and the firing threshold of that same neuron as $v_{th,k}$. For $W_k=1$, we aim to push the membrane potential above the threshold, while for $W_k=0$, the goal is to maintain it below it.

The straightforward option would be to directly train the binary spike produced by the neuron. However, the spike only indicates whether the threshold was crossed, and because the firing operation is non-differentiable, directly supervising the spike does not provide enough information about how far the current membrane state is from the desired value. To avoid this, we train with the membrane potential just before the spike decision and reset.

More precisely, we use the signed distance between the membrane potential and the firing threshold,
\begin{equation}
m_k=\frac{v_k-v_{th,k}}{\tau_{wm}},
\label{eq:margin}
\end{equation}
where $\tau_{wm}$ controls the scale of the watermark loss. A positive value indicates that the selected neuron's membrane potential is above the threshold, while a negative value means it is below it. 
We then optimize
\begin{equation}
\mathcal{L}_{wm}=\frac{1}{B}\sum_{k=1}^{B}
\mathrm{BCE}(\sigma(m_k),W_k).
\label{eq:lwm}
\end{equation}
For a one-bit, the loss pushes $v_k$ above $v_{th,k}$, while for a zero-bit, it pushes the membrane potential in the opposite direction. This way, the optimizer can receive information about the distance to the threshold even when the binary spike has not changed yet. Section~\ref{sec:transfer} compares our approach with direct spike supervision.

To limit the changes the watermarking produces in the protected model, we keep a frozen copy $M_0$ of the original checkpoint. For a task batch $x$, we optimize
\begin{equation}
\mathcal{L}=\mathcal{L}_{task}
+\lambda_{fid}\,\mathrm{KL}\!\left(M_0(x)\,\|\,M(x)\right)
+\lambda_{wm}\mathcal{L}_{wm}.
\label{eq:total}
\end{equation}

The first term corresponds to the original task, the second keeps the protected checkpoints' behavior close to $M_0$, and the last term embeds the watermark. The additional training settings are reported in Appendix~\ref{app:pareto}.

\subsection{Watermark Verification}
\label{sec:extraction}

After the protected model is released, the verifier runs the secret input through the suspect checkpoint and reads the coordinates stored in $C$. We consider two ways of reading the membrane watermark: a binary read based on the firing threshold and a continuous read based on the distance to the same.

\textbf{Hard membrane read.} The simplest option is to recover each bit according to which side of the firing threshold the selected state is located.
\begin{equation}
\hat W_k=\mathds{1}[v_k-v_{th,k}\geq0].
\label{eq:hardread}
\end{equation}

Thus, a membrane potential above the threshold recovers a 1, while a value below it recovers a 0, equivalent to the emitted spikes of the neuron expressed in Eq.~\eqref {eq:fire}. Although we use the continuous membrane state during training, the final binary read uses the same threshold decision that the LIF neuron already uses. Thus, the verifier does not require an additional learned decoder to recover the payload.

We measure the number of incorrectly recovered bits using the bit error rate (BER):
\begin{equation}
\mathrm{BER}=\frac{1}{B}\sum_{k=1}^{B}\mathds{1}[\hat W_k\neq W_k].
\label{eq:ber}
\end{equation}

A BER of 0 means that all payload bits are correctly recovered. We also report the number of matching bits to make the ownership rule easier to interpret (e.g., $48$ matching bits out of a $64$-bit payload).

\textbf{Membrane-distance score.} The hard read only checks whether the threshold is crossed, without considering the distance to the threshold itself. To keep this information, we also calculate a continuous membrane score in the pre-fire membrane:
\begin{equation}
q_k=\sigma\!\left(a_k\frac{v_k-v_{th,k}}{\tau_{wm}}\right),
\qquad
S_{mem}=\frac{1}{B}\sum_{k=1}^{B}q_k.
\label{eq:softscore}
\end{equation}
Here, $a_k=2W_k-1$ indicates the expected side of the threshold the potential should be in. A value of $q_k$ close to 1 means that the membrane state is far on the expected side of the threshold, while a value close to 0 means that it is far on the wrong side, and values around $0.5$ correspond to states close to the threshold.
We use the hard membrane read as the main ownership rule and report $S_{mem}$ as an additional MeMark verification variant. In Section~\ref{sec:nullcalib} we evaluate both methods on genuine and unrelated keys.

\subsection{Output Watermark}

In addition to the internal membrane watermark, MeMark can also embed a response in the output of the model. This variant is useful when the verifier cannot inspect the internal LIF states and has only query access to the model. It uses the same secret challenge but stores a separate payload in the output response.

When the models expose the output scores, we train pairwise score margins corresponding to the output bits by choosing two possible outputs and comparing their respective scores. For classification or text-generation interfaces, we represent the response using the chosen labels or tokens. The output payload is decoded independently from the membrane payload. Additional details on the output response and its decoding are given in Appendix~\ref{app:decoding}.

The output watermark can be verified without white-box access, but it depends on the output interface. Thus, if the output head is replaced, this response can be removed while leaving the internal membrane watermark unchanged. The membrane watermark is intended to survive checkpoint reuse, while the output watermark provides an additional verification option when only the model output is available.
\section{Evaluation Setup}
\label{sec:setup}

We evaluate MeMark on eight SNN testbeds covering different model sizes, architectures, and tasks. We use the smaller recurrent SNN for the repeated ablations and attack experiments, while SpikeGPT~\cite{spikegpt} is used to test the watermark on a more realistic, substantially larger pretrained checkpoint. We also evaluate six vision models to showcase that the method transfers across convolutional, residual, and transformer-based SNNs. The complete evaluation setup is reported in Appendix~\ref{app:testbeds}.

\subsection{Testbeds}

The recurrent model is a 4-layer SNN with 0.97M parameters trained on enwik8~\cite{mahoney2011enwik8}, while SpikeGPT-216M~\cite{spikegpt} contains 215.4M parameters and is pretrained on OpenWebText~\cite{gokaslan2019openwebtext} (for the watermark embedding, we use enwik8 as described below). For the vision testbeds, we consider spiking CNNs, SEW-ResNet-18~\cite{sewresnet}, Spikformer~\cite{spikformer}, and QKFormer~\cite{qkformer} for both traditional datasets like MNIST and CIFAR-10 and neuromorphic ones like CIFAR10-DVS~\cite{cifar10dvs} and N-Caltech101~\cite{ncaltech101}.


\subsection{Watermark Training}

For the recurrent model, we use a 32-bit payload in most experiments to increase the number of ablations and repeated attacks. As shown in Appendix~\ref{app:ablations}, the model can recover payloads from 8 to 128 bits using the same number of training steps, so the payload length decision has little effect on the original task performance.

For SpikeGPT, we use a bigger 64-bit payload and embed 20 independent watermark keys in total. We use the first 5 of these 20 keys to select the embedding budget and keep the remaining 15 out of this selection. These keys are trained for $3\,000$ steps and evaluated at $1\,200$, $2\,000$ and $3\,000$ steps using the calibration population described below. The genuine responses stop improving after $2\,000$ embedding steps for all SpikeGPT tested keys. The complete budget comparison is reported in Appendix~\ref{app:budget}.

\subsection{Key Verification}

To set the threshold used for verification and evaluate its resistance to random keys, we use two separate sets of keys. The first set is used only to choose the threshold, while the second one is kept until the final evaluation. For SpikeGPT, we use $6\,000$ random keys for the initial threshold selection and five genuine keys evaluated on checkpoints trained with a different key and on a clean non-watermarked checkpoint. The $6\,000$ random keys are each tested on the five protected models and the clean model, giving $36\,000$ negative claims. Each genuine key is then tested on the four protected models in which it was not embedded and on the clean model, adding another 25 claims. In total, this gives $6\,005$ key identities and $36\,025$ negative model-key claims for calibration.

After the threshold is fixed, we evaluate $30\,000$ new random keys on all 20 protected checkpoints and the clean checkpoint. Together with the cross-key controls, this gives $630\,055$ negative claims from $30\,020$ key identities. We measure false positives per random key. A key is counted as a false positive if any of the 21 checkpoints accept it, since counting all model-key claims separately would dilute a false acceptance across the 21 checkpoints.

\section{Metrics}

We evaluate the watermark's success using the BER as described in Eq.~\eqref{eq:ber}, which measures the fraction of incorrectly recovered watermark bits. We also report the membrane score $S_{mem}$ from Eq.~\eqref{eq:softscore}, where larger values indicate that the selected membrane potentials are farther on the expected side of their firing thresholds.

To evaluate if the watermark is specific to the secret input, we measure the off-trigger agreement, where a value close to $0.5$ indicates that the non-challenge input does not reveal the payload and thus is not related to the model. Finally, we also measure the original task performance before and after watermarking to see the effect the watermark has on the model's behavior, using validation loss for the language models and clean accuracy for the vision tasks.

For the recurrent two-channel experiments, the joint score is $\min(m_{\mathrm{mem}},m_{\mathrm{out}})$. We set its threshold one bit above the largest non-matching calibration score, giving $\geq22/32$. This joint rule is used only when both channels are evaluated, while the hard membrane read remains our primary rule.
\section{Experimental Results}
\label{sec:results}

\subsection{Watermark Verification}
\label{sec:nullcalib}

We first use the first 5 SpikeGPT keys to select the number of embedding steps. We train them up to $3\,000$ steps and evaluate the watermark at $1\,200$, $2\,000$, and $3\,000$ steps. At $1\,200$ steps, the weakest genuine key recovers 53/64 bits, while at $2\,000$ steps this increases to 59/64. Continuing the training to $3\,000$ steps does not improve this result, as the weakest key still recovers 59/64 bits. We therefore use $2\,000$ embedding steps for the SpikeGPT experiments. The complete comparison is reported in Appendix~\ref{app:budget}.

We use the same experiment to select the verification threshold. The strongest negative key observed during the calibration reaches 50/64 bits, so based on this result we set the verification threshold to 51 matching bits. This value is fixed before evaluating the remaining 15 genuine keys and the 30000 fresh random keys.

\begin{table}[t]
\centering
\footnotesize
\setlength{\tabcolsep}{3.0pt}
\caption{SpikeGPT ownership decisions. The fixed 51/64 rule accepts all 5 calibration and 15 held-out genuine keys, giving 20/20 overall, while none of the 30000 fresh random keys pass across all 20 protected checkpoints and the clean model.}
\label{tab:ownership}
\begin{tabular}{@{}lcccc@{}}
\toprule
Fixed rule & FP keys & Key FPR & 95\% upper & Genuine \\
\midrule
Hard/spike, $\geq51/64$ &
0/30\,000 &
0.000\% &
0.010\% &
20/20 \\
\bottomrule
\end{tabular}
\end{table}

We next evaluate all 20 independent 64-bit keys using the selected $2\,000$-step budget. The first 5 are the calibration keys above, while the remaining 15 were not used to select the budget or threshold. All 20 pass the fixed threshold, recovering between 55 and 62 bits, with an average of 59.4.

Then, we test the selected threshold on $30\,000$ fresh random keys across all 20 protected checkpoints and the clean checkpoint. None of the random keys pass the 51/64 rule, giving a false-positive rate of $0.000\%$ and a one-sided 95\% upper bound of $0.010\%$ (Table~\ref{tab:ownership}). In Appendix~\ref{app:budget}, we also compare this result with shorter embedding step checkpoints.

We also evaluate the continuous membrane score $S_{mem}$. Using the calibrated cutoff of $0.738$, 2 of the $30\,005$ negative and cross-key identities pass while all 20 genuine keys are accepted. Therefore, the continuous score can also separate the genuine and unrelated keys with a low false-positive rate. However, as shown later in Section~\ref{sec:adaptive}, an attack that directly reduces the membrane margins affects this score more than the hard read. For this reason, we keep the hard threshold as the main ownership rule.

Finally, we measure whether the longer embedding affects the original model. At $2\,000$ steps, the mean validation loss of the 20 watermarked checkpoints is $5.0506$, compared with $5.0551$ for the equal-budget anchored controls. The difference is only $-0.0046$, showing that the additional embedding steps do not introduce a measurable loss from the watermark itself.

On the smaller recurrent SNN, all 10 genuine keys recover perfectly within the fixed training budget, and those same keys on unwatermarked models remain close to random agreement (between $0.469$ and $0.562$).

\subsection{Fitted Ownership Claims}
\label{sec:falseclaim}

The previous experiments assume that the keys are generated independently of the model. However, a dishonest claimant with white-box access can instead inspect a model and then choose a key that already matches its internal states. We refer to this scenario as a fitted claim.

We test this attack on the recurrent model without modifying its weights. A claimant who optimizes only the trigger without having control over the selected coordinates and bits can recover 18/32 membrane bits, which remains below the verification threshold. However, if the claimant is allowed to inspect the model before choosing  the coordinates and bit values, the same checkpoint immediately reaches 32/32 bits, as it can select already occurring states under observed inputs. We observe a similar response on the output response, starting at 16/32, but selecting token pairs that already match the model's output preferences raises it to 32/32 (Table~\ref{tab:falseclaim}).

\begin{table}[t]
\centering
\footnotesize
\caption{False claims fitted to an unwatermarked model and evaluated at the calibrated joint threshold of $\geq22/32$ (Accepted claims are shown in bold). A claimant who chooses the key after inspecting the model can make the model-level test accept even though no watermark was embedded.}
\label{tab:falseclaim}
\begin{tabular}{@{}lccc@{}}
\toprule
Accuser's strategy & Mem. & Aux. out. & Joint \\
\midrule
Honest random key & 13/32 & 20/32 & 13/32 \\
Coordinates and bits fitted & \textbf{32/32} & 16/32 & 16/32 \\
Trigger optimized only & 18/32 & 20/32 & 18/32 \\
Coordinates fitted $+$ trigger optimized & \textbf{32/32} & 17/32 & 17/32 \\
\quad $+$ output token pairs fitted & \textbf{32/32} & \textbf{32/32} & \textbf{32/32} \\
\bottomrule
\end{tabular}
\end{table}

These fitted claims support the requirement for the pre-release commitment described in Eq.~\eqref{eq:commitment}. The watermark response by itself only shows that a model matches a key. It does not show whether that key was embedded before release or selected later after inspecting the model. We observe the same problem with the continuous score, as fitting the coordinates and bits raises $S_{mem}$ to $1.000$, above the scores obtained by the genuine keys. Thus, the owner must open the commitment created before publication and verify the same registered key on the suspect model.

However, registering a large number of keys can still increase the chance that one of them later matches another model. We discuss this case, together with competing ownership claims, in Appendix~\ref{app:ambiguity}.

\subsection{Membrane Watermark}
\label{sec:transfer}

We first evaluate why MeMark is trained with the membrane distance instead of directly training the binary spike. When we supervise the spike output itself, the payload remains close to random with BER around $0.5$. In contrast, when we use the pre-fire membrane-margin loss, BER drops to $0.00$. The main difference is that the membrane value tells the optimizer how far the neuron is from the desired side of the threshold, even when the spike has not changed yet. With the binary spike, two membrane values on the same side of the threshold look identical, while the membrane distance still provides a direction for training.

After embedding, however, the continuous value is not required to recover the payload, as the hard read from Eq.~\eqref{eq:hardread} and the actual LIF spike recover exactly the same bits on the 20 SpikeGPT checkpoints and on all $630\,055$ negative claims. Thus, the membrane distance is useful for training, while the neuron's own threshold decision is enough for the final binary read. We keep $S_{mem}$ as an additional score for the distance to this boundary.

We also compare MeMark with a DICTION-style activation watermark adapted to the same SNN setting. DICTION~\cite{diction} stores the watermark in the internal activations and uses a learned projection to recover the payload. For a controlled comparison, we use the same challenge, payload, checkpoint, optimizer, fidelity term, output objective, and training budget for both methods. The only differences are the internal carrier and how the watermark is decoded. Both methods recover all 32 bits and have a similar effect on the validation loss, with an increase of $+0.223$ for MeMark and $+0.238$ for DICTION. Thus, both methods perform similarly when considering direct watermark recovery. However, MeMark does not require an additional learned projection, as the LIF firing threshold already provides the boundary used to read each bit. The difference between the two methods also becomes more noticeable when the secret challenge is not present, where MeMark reveals less information about the payload, as shown in Section~\ref{sec:selectivity}.

\subsubsection{Challenge-Response Selectivity}
\label{sec:selectivity}

We next investigate what happens when the secret challenge input is not used. Ideally, the watermark should only reveal the secret payload under the registered challenge and should not be recoverable from normal inputs. We test the same watermarked models using random tokens and held-out text. Furthermore, we additionally test wrong challenge inputs and clean models as additional controls.
\begin{figure*}[!t]
\centering
\begin{subfigure}{0.49\textwidth}\centering
\includegraphics[width=0.8\linewidth,trim=0pt 7pt 0pt 8pt, clip]{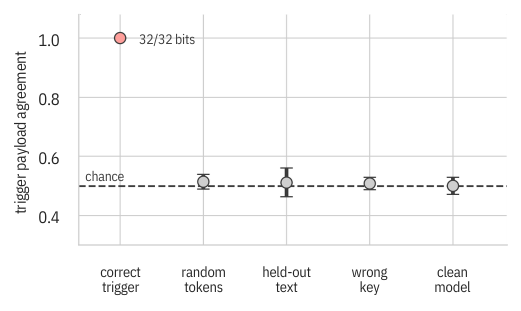}
\caption{Only the correct challenge reveals the payload. Mean $\pm$ std over ten keys.}
\label{fig:selconditions}
\end{subfigure}\hfill
\begin{subfigure}{0.49\textwidth}\centering
\includegraphics[width=0.8\linewidth,trim=0pt 7pt 0pt 8pt, clip]{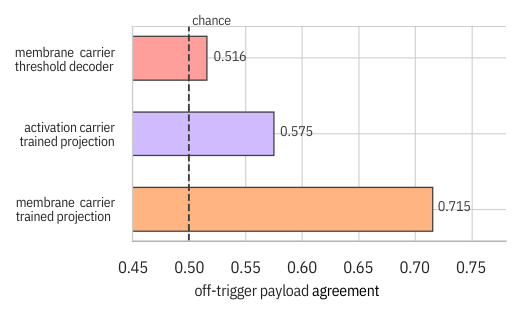}
\caption{Off-trigger leakage on different payload reading methods.}

\label{fig:seldecoder}
\end{subfigure}
\caption{Challenge-response selectivity. Both panels use off-trigger payload agreement: the mean agreement on the key's one-bits and zero-bits. A value of $0.5$ indicates no payload information.}
\label{fig:selectivity}
\end{figure*}

As shown in Figure~\ref{fig:selectivity}, MeMark recovers all 32 bits with the correct challenge, and when the input challenge is removed, the mean off-trigger agreement over ten keys is $0.516$, close to the random value of $0.5$. The matched DICTION-style watermark also recovers all bits with the correct challenge, but its off-trigger agreement increases to $0.575$. Thus, the difference between both methods is small when looking only at watermark recovery, but MeMark reveals less of the payload when the secret challenge is not present.

We next investigate whether this difference comes from the membrane values themselves or from how they are decoded. For that, we keep the same membrane vector and replace the firing-threshold read of MeMark with a trained projection. With the learned projection decoder, the agreement on non-challenge inputs increases from  $0.516$ to $0.715$.We also test activation-based decoders that use zero, the mean, or the median activation (measured on calibration inputs) as the reference for reading each bit. All of them reveal more of the payload on unrelated inputs than MeMark, as shown in Table~\ref{tab:decodercontrols}. These results suggest that using the neuron's own firing threshold helps keep the payload from being recovered on unrelated inputs.

\begin{table}[t]
\centering
\footnotesize
\setlength{\tabcolsep}{2.2pt}
\caption{Carrier and decoder controls over ten keys. Off-trigger agreement is measured on 300 probes per key; $0.5$ is the chance level.}
\label{tab:decodercontrols}
\begin{tabular}{@{}llcc@{}}
\toprule
Carrier & Decoder reference & On-trig. BER & Off-trig. agreement \\
\midrule
membrane & \textbf{firing threshold} & 0.000 & \textbf{0.516} \\
membrane & trained projection & 0.000 & 0.715 \\
activation & zero & 0.000 & 0.580 \\
activation & calibration mean & 0.000 & 0.680 \\
activation & calibration median & 0.000 & 0.677 \\
\bottomrule
\end{tabular}
\end{table}

Looking at the predictions explains part of this difference: with unrelated inputs, only $4.8\%$ of the threshold reads are one, while the trained projection predicts one at $48.3\%$ of the positions, making it easier to reconstruct part of the payload by chance. We also observe that the mean and median activation references move by $0.28$ and $0.26$ per coordinate after 100 task-only steps. In contrast, the firing threshold is already part of the neuron and does not require storing or recalibrating an additional reference value.

\subsubsection{SNN-Specific Attacks}
\label{sec:carriermatrix}

Since the firing threshold is also part of the LIF dynamics, we test whether changing the neuron parameters affects the watermark. We modify the firing threshold $v_{th}$ and the membrane time constant $\tau$, which control when the neuron spikes and how its membrane state evolves over time. Moderate changes to either parameter do not affect MeMark or the matched DICTION-style watermark. When the changes become larger, both methods eventually degrade, although in different ways: lowering $v_{th}$ affects the activation projection more, while increasing $\tau$ has a larger effect on the membrane carrier. Importantly, the settings that produce a large BER also substantially increase the task loss, meaning that the model behavior is already being affected. The complete sweep is reported in Appendix~\ref{app:perturbcost}.

\subsection{Comparison with Previous Watermarks}
\label{sec:baselines}
We next compare MeMark with previous white-box and SNN watermarking methods on the recurrent model. Since these approaches store ownership evidence in different parts of the network, not all methods provide the same verification interfaces. As such, we start from the same checkpoint and use the same payload size, training budget, and objective whenever possible.

Poursiami et al.~\cite{poursiami} store a continuation in the model output, while SpikeTimer~\cite{spiketimer} is designed for model authorization. For the internal watermark, Table~\ref{tab:baselinenumbers} places the matched DICTION result evaluated above alongside DeepSigns~\cite{deepsigns} and Uchida~\cite{uchida}. DICTION and DeepSigns use internal activations, while Uchida embeds the watermark directly in the model weights.

\begin{table*}[!t]
\centering
\footnotesize
\setlength{\tabcolsep}{2.5pt}
\caption{Baselines on the 4-layer recurrent SNN, reimplemented from the same checkpoint and training budget where the method allows it. Trigger-conditioned methods are also read on off-trigger probes; DeepSigns is evaluated on held-out data and Uchida directly from weights. Fidelity is the resulting loss after the defense is applied.}
\label{tab:baselinenumbers}
\begin{tabular}{@{}llllll@{}}
\toprule
Scheme & Payload & White-box read on the challenge & Read without the challenge & Output channel & Fidelity \\
\midrule
\textbf{MeMark} (ours) & 32 bits & BER $0.000$ & agreement $0.516$ & BER $0.000$ & loss $2.413$ \\
DICTION-style~\cite{diction} & 32 bits & BER $0.000$ & agreement $0.575$ & BER $0.000$ & loss $2.428$ \\
DeepSigns-style~\cite{deepsigns} & 32 bits & BER $0.000$ & BER $0.125$ & n/a & loss $2.447$ \\
Uchida et al.~\cite{uchida} & 32 bits & BER $0.094$ & no challenge to withhold & n/a & loss $2.413$ \\
Poursiami et al.~\cite{poursiami} & 1 continuation & n/a & n/a & 16/16 tokens & loss $2.284$ \\
\bottomrule
\end{tabular}
\end{table*}

Overall, MeMark achieves exact recovery of the 32-bit internal payload while keeping the verification tied to the secret challenge. Compared with the matched projection-based decoder baseline, MeMark has the same on-challenge recovery and similar recovery cost, but it does not need the $917\,504$-weight learned projection, and has lower off-trigger payload agreement. Head replacement also separates MeMark from current output-based SNN watermarks, as our primary response remains in the backbone even after the original output interface is removed.

\subsection{Robustness to Model Modifications}
\label{sec:checkpointmods}

Next, we evaluate modifications that are more likely to occur when a released checkpoint is reused. We consider replacing the output head, fine-tuning the model, pruning its weights, and quantizing it. These experiments are particularly useful for comparing the different MeMark carriers, as some modifications affect the model output directly, while others also change the internal representation.

We first reset and retrain the output head. As expected, this removes the output watermark on the recurrent SNN, since the response is stored in the part of the model being replaced. The membrane watermark is different. When the backbone is kept frozen, its BER remains $0.00$. If we also fine-tune the backbone for a new task, the membrane response changes during the first updates but returns to BER $0.00$ by step 200. We observe the same separation on SpikeGPT. Across all 20 keys, the membrane read remains at its released value after head replacement, while the output BER increases to $0.507$ (see Table~\ref{tab:robustreal}). Thus, the internal watermark remains available even when the original output head is no longer present.

Pruning and quantization produce a similar result for the membrane carrier. Neither $90\%$ pruning nor int8 quantization changes any membrane bit for the 20 SpikeGPT keys. Fine-tuning has a larger effect because it directly updates the backbone parameters. After 200 steps, the models lose 1.1 matching bits on average, but all 20 keys still satisfy the 51/64 verification threshold. The weakest response after fine-tuning is 53 bits.

We obtain a weaker result with the shorter training budgets. When all keys are trained for $1\,200$ steps, 19 of the 20 still pass after fine-tuning, as one key drops from 53 to 45 bits. At $2\,000$ steps, this same key reaches 59 bits before fine-tuning and 56 afterward. However, the initial number of matching bits does not completely determine how much a key will change, as some keys gain bits while others lose them. Thus, a larger margin gives the watermark more room to change before reaching the threshold, but it does not guarantee that every key will be affected in the same way. The complete comparison is reported in Appendix~\ref{app:budget}.

\begin{table}[t]
\centering
\footnotesize
\caption{Attacks on all 20 SpikeGPT checkpoints after $2\,000$ embedding steps. Matching bits are the mean over keys, with the observed range. ``Pass'' uses the fixed 51/64 rule.}
\label{tab:robustreal}
\begin{tabular}{@{}lcc@{}}
\toprule
Attack & Matching bits & Pass \\
\midrule
None (baseline) &  59.4 [55, 62] &  20/20 \\
Fine-tune, 200 steps &  58.3 [53, 61] &  20/20 \\
Prune $90\%$ &  59.4 [55, 62] &  20/20 \\
int8 quantization &  59.4 [55, 62] &  20/20 \\
Head reset $+$ retrain &  59.4 [55, 62] &  20/20 \\
Channel permutation &  32.6 [28, 38] &  0/20 \\
\quad after realignment &  59.4 [55, 62] &  20/20 \\
\bottomrule
\end{tabular}
\end{table}

The pruning result also shows an important trade-off for an attacker. At $90\%$ pruning, all 20 membrane watermarks still verify, but the validation loss increases from about $5.05$ to $46.2$. In this case, pruning damages the language model much more than the ownership signal.

We further test these modifications on the recurrent SNN using longer sweeps. Across four independently watermarked models, both watermark channels remain at BER $0.00$ through $90\%$ pruning and int8 quantization. The membrane response is also unchanged after 400 task fine-tuning steps, and ANN--SNN reconversion preserves both responses. Even after combining $90\%$ pruning with 250 fine-tuning steps, the membrane BER remains $0.00$ and the output BER is at most $0.031$.

Besides modifying the model, an SNN user may also change how temporal inputs are processed. We therefore add timestamp jitter to CIFAR10-DVS, up to $\sigma=50\,000$. At this value, $22\%$ of the integrated input tensor changes, but both the membrane watermark and the output BER remain at their original BER of $0.00$. This experiment measures tolerance to timing noise rather than a watermark-aware attack. We consider the watermark-aware attack separately in Appendix~\ref{app:eventretiming}.

Finally, we test what happens when part of the key becomes known. We reveal the trigger and a subset of the coordinate-bit pairs, simulating an attacker that obtains part of the secret key. The disclosed coordinates can be inverted, effectively eliminating their payload, but the remaining secret coordinates stay at BER $0.00$, and the separately keyed output response remains unchanged. This behavior is expected, as knowing one part of the sparse carrier does not reveal where the remaining bits are stored. The complete partial-exposure experiment is reported in Appendix~\ref{app:partialknowledge}.

\subsection{Structural Modifications}
\label{sec:permutation}

The previous experiments change model parameters but keep the internal channel ordering intact. However, a checkpoint can also be modified without changing its function. For example, hidden channels can be permuted as long as the inverse permutation is applied to the following layer. The model then produces the same output, but the numerical indices stored in the watermark key no longer point to the same neurons, so the key is no longer verifiable even though the individual neurons are unchanged.

We first test this case on the recurrent SNN by permuting all 128 channels and compensating for the change in the next layer. The maximum difference in the logits is only $2.9\times10^{-6}$, and the output watermark remains exact, confirming that the model behavior is effectively unchanged. Directly reading the original membrane indices increases BER to $0.438$. However, the watermark has not been removed. Instead, the neurons carrying it have moved to different positions.

\begin{figure}[t]
\centering
\includegraphics[width=0.8\linewidth,trim=0pt 8pt 0pt 5pt, clip]{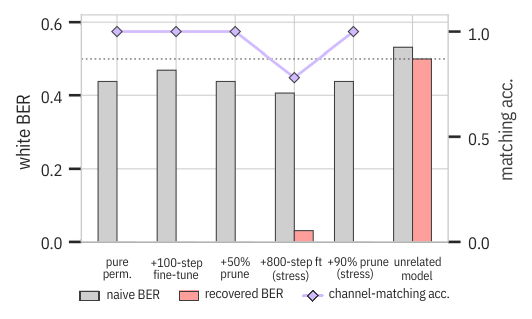}
\caption{Model accuracy and BER with function-preserving channel permutations before and after recovering the coordinates by matching on non-secret signatures against a previous protected checkpoint.}
\label{fig:permdefense}
\end{figure}

To recover the moved coordinates, we compare the channel responses of the suspect model with those of the protected release on a small set of non-secret inputs. We then pair channels with similar responses and translate the watermark coordinates to their new positions. The secret challenge and payload are not used during this process. Matching the 128 channels takes about 13 seconds and restores all keyed coordinates, as shown in Figure~\ref{fig:permdefense}.

We next test whether this recovery still works after the suspect model has also been modified. After combining the permutation with 100 fine-tuning steps, $50\%$ and $90\%$ pruning, and an 800-step stress test, the recovered BER remains at or below $0.031$. We additionally repeat the experiment with five new random permutations and three independently trained reference/suspect pairs. In contrast, aligning an unrelated model leaves BER at $0.531$. This control is important because it shows that the alignment is recovering corresponding channels from their behavior rather than searching for a permutation that happens to match the secret payload.

A permutation keeps the number of channels fixed, so we also test a transformation that changes the network width. Each channel is duplicated, its outgoing weights are divided between both copies, and the resulting 256 channels are then permuted. The transformed network remains numerically equivalent to the original, but direct verification gives BER $0.562$. Matching the 128 reference channels against the 256 channels of the suspect model still restores BER $0.00$, while the unrelated control again remains at BER $0.531$.

\begin{table}[t]
\centering
\footnotesize
\caption{Function-preserving structural transformations. ``Recovered'' uses the retained protected checkpoint to align channels before applying the secret key.}
\label{tab:structural}
\begin{tabular}{@{}lccc@{}}
\toprule
Transformation & Max logit $\Delta$ & Naive BER & Recovered BER \\
\midrule
Layer split ($W_2W_1$) & $0$ & 0.000 & 0.000 \\
Replicate $+$ permute & $7{\times}10^{-6}$ & 0.562 & \textbf{0.000} \\
\quad $+$ layer split & $7{\times}10^{-6}$ & 0.562 & \textbf{0.000} \\
\bottomrule
\end{tabular}
\end{table}

It is also important to note that not every reparameterization used for ANNs remains function-preserving in an SNN. Scaling a layer without changing the LIF threshold also changes when neurons spike, moving the logits by as much as $5.1$ in our experiments. When the threshold is scaled together with the layer, both the model function and the watermark are preserved, as MeMark's decoder is part of the natural firing behavior on the neurons. We consider a stronger attack against the structural recovery in Appendix~\ref{app:structuralfull}.

\subsection{Different SNN Architectures}
\label{sec:architecture}

We test whether the same membrane watermark can also be embedded in other SNN architectures. We use six vision testbeds covering convolutional and residual SNNs, as well as two transformer architectures, Spikformer and the hierarchical QKFormer.

\begin{figure*}[ht]
\centering
\begin{subfigure}{0.49\textwidth}\centering
\includegraphics[width=0.8\linewidth,trim=0pt 8pt 0pt 10pt, clip]{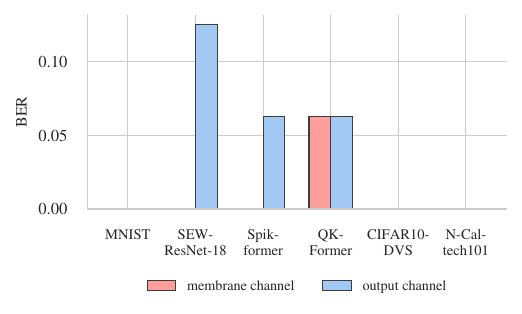}
\caption{Watermark recovery. Bars are absent where BER is exactly $0.000$.}
\label{fig:visionber}
\end{subfigure}\hfill
\begin{subfigure}{0.49\textwidth}\centering
\includegraphics[width=0.8\linewidth,trim=0pt 8pt 0pt 8pt, clip]{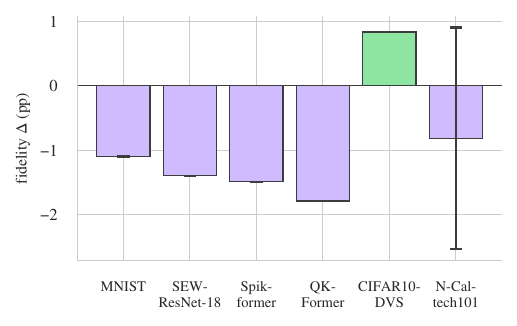}
\caption{Fidelity relative to an equal-budget task-only control.}
\label{fig:visionfid}
\end{subfigure}
\caption{Vision feasibility across six testbeds: three different SNN architectures on CIFAR-10 and three additional datasets evaluated with a shared spiking CNN.}
\label{fig:visionbar}
\end{figure*}

The results in Figure~\ref{fig:visionbar} show that five of the six testbeds recover the complete membrane payload with BER $0.00$. QKFormer is the only exception, with BER $0.0625$, corresponding to one incorrect bit in the 16-bit payload. The optional output watermark remains below BER $0.15$ in all cases. We also compare the protected models with their equal-budget controls. On MNIST, SEW-ResNet-18, CIFAR10-DVS, and N-Caltech101, the difference remains below 1.5 accuracy points. These results show that the membrane carrier is not limited to the recurrent language models used in the main experiments.

The transformer models require more training than the convolutional models. With $3\,000$ embedding steps and a weaker KL term, Spikformer reaches membrane BER $0.00$ with a $1.49$-point accuracy difference from its control. While QKFormer reaches BER $0.0625$ with a $1.79$-point difference. Thus, the same watermark mechanism can be used across these architectures, but the embedding settings still need to be adjusted for the model being protected. The training settings for these models are reported in Appendix~\ref{app:pareto}.

\subsection{Verification Cost}
\label{sec:cost}

Besides robustness, the watermark should also be practical and computationally cheap to verify. For the optional output response, a single language-model query recovers 24/32 designated bits. Repeating the query with the decoding settings from Appendix~\ref{app:decoding} recovers between 22 and 24 bits, while the wrong-key controls recover no bits.

For the membrane watermark, the verifier only needs to observe the states selected by the key. A 256-token SpikeGPT input produces $1\,769\,472$ membrane values, but a 64-bit key uses only 64 of them, corresponding to $0.0036\%$ of the available states. These coordinates are distributed across 30 of the 36 LIF modules, so the verifier does not need to record the complete internal state of the model. This suggests that the latency could be further reduced by limiting the number of modules targeted by the watermark. However, as shown in Appendix~\ref{sec:ablations}, placing watermark bits too close together on the same neurons can make them harder to embed. Thus, using fewer modules introduces a trade-off between verification cost and how freely the watermark coordinates can be selected. The key itself is also small, requiring about $1.6$ KB compared with the $861.7$ MB checkpoint. In comparison, the learned $32\times28\,672$ projection used by the matched DICTION-style baseline contains $917\,504$ weights, or about $3.5$ MB. MeMark does not require this additional decoder, since the firing thresholds used to read the bits are already part of the model. As a result, the information that MeMark needs to store for verification is about $\times2\,300$ smaller than the learned projection. The complete verification-cost measurements are reported in Appendix~\ref{app:verificationcost}.

\subsection{Adaptive Attacks}
\label{sec:adaptive}

The previous experiments consider modifications that may affect the watermark as a side effect of reusing the checkpoint. We now consider a stronger attacker who knows how MeMark works and deliberately tries to remove it, but does not know the secret key.

We first test whether the attacker can identify the membrane coordinates selected by the key. Since the coordinates are secret, the attacker searches for states that look unusual: states that sit far from the firing threshold, remain stable across different inputs, or the ones that can be easily moved by changing the input. We rank all $32\,768$ candidate states using these properties and take the highest-ranked ones as guesses for the watermark coordinates. The attacker then optimizes a synthetic trigger for these guessed states. If the watermark coordinates were easy to distinguish from normal membrane states, we would expect them to appear near the top of this ranking. However, none of the first 64 candidates belongs to the real key, and only one keyed coordinate appears among the first 512. The synthetic trigger also shares no tokens with the real challenge and recovers only 16/32 bits for the top-64 search and 19/32 for the top-512 search. After 400 removal steps, both the membrane and output responses of the owner remain at BER $0.00$. Thus, this search is not sufficient to reliably identify the watermark coordinates.

An attacker may instead try to look for unusual behavior that distinguishes a watermarked model from a clean one. We first test this using an activation-outlier detector, which checks whether the secret challenge produces internal activations that stand out from those produced by normal inputs. With this detector, the challenge is indeed flagged as an outlier, but the same input is also flagged on a clean model. Therefore, the detector responds to the unusual input rather than the watermark itself. We also use a Neural-Cleanse-style~\cite{neuralcleanse} reconstruction that looks for positions that can be driven unusually easily by a reconstructed input. However, only one of the ten keyed positions is detected, while the other 9 are indistinguishable from the 12 clean controls. Additional detection experiments are reported in Appendix~\ref{app:detection}.

An attacker who does not know which states are selected can still try to weaken all of them at once. At every update, we sample fresh random tokens and collect all $32\,768$ pre-fire membrane states. The attacker then pushes every state toward its firing threshold, reducing the margins that could be carrying watermark bits. At the same time, the normal task and KL losses are kept during training so that the model remains useful. We vary the strength of the new objective from $\lambda=0.01$ to $1\,000$ and find that $\lambda=100$ produces the largest watermark change without a large increase in task loss.

Unlike the previous attacks, this one does weaken the watermark. After 250 steps, the mean membrane BER over four keys increases from $0.000$ to $0.0625$, while the output BER reaches $0.031$. The held-out task loss increases by only $0.037$, so this change is not simply caused by destroying the model. However, the remaining watermark is still strong enough for all four models to satisfy the fixed 22/32 joint verification rule. Thus, attacking every membrane state is more effective than trying to guess the secret coordinates, but it is only able to weaken it rather than remove the ownership evidence. This attack affects the continuous score more directly. At $\lambda=1\,000$, the hard read loses $3.1\%$ of its range, while $S_{mem}$ loses $18.4\%$.

\subsection{Rollback Attack}
\label{sec:rollbackresults}

The adaptive attacks above work directly on the released model without another reference. A different alternative is to compare the protected checkpoint with a related checkpoint and try to identify which parameters changed during embedding. We evaluate this with a rollback attack that ranks parameters taking into account the difference between both models and restores the largest changes (we refer to the fraction of parameters restored as the rollback budget). The quality of the reference matters: the closer it is to the model immediately before watermarking, the easier it should be to isolate the watermark updates.

To study this effect, we start from a public model and privately adapt it for 400 steps to obtain $M_{prop}$. We then embed MeMark and release only the protected checkpoint. The attacker receives three references that could realistically be available: the original public model, another model adapted on the same corpus, and a model adapted on a different corpus. We additionally test the exact private $M_{prop}$ to measure the stronger case where the checkpoint immediately before watermarking is also available.

\begin{table}[t]
\centering
\footnotesize
\setlength{\tabcolsep}{2.5pt}
\caption{Differencing attack on private proprietary pre-embedding checkpoint, mean over five keys. Before attack, the released models read $0.119/0.000$ at clean loss $4.428$.}
\label{tab:lifecycle}
\begin{tabular}{@{}lcccc@{}}
\toprule
Reference & Revert & Mem.\ BER & Out.\ BER & Clean loss \\
\midrule
\multirow{3}{*}{Public initializer}
 & $0.1\%$ & 0.156 & 0.019 & 4.496 \\
 & $1\%$   & 0.169 & 0.094 & 4.593 \\
 & $5\%$   & 0.206 & 0.256 & 4.914 \\
\midrule
\multirow{2}{*}{Similar, same corpus}
 & $0.1\%$ & 0.169 & 0.019 & 4.427 \\
 & $1\%$   & 0.244 & 0.119 & 4.425 \\
\midrule
\multirow{3}{*}{Similar, disjoint corpus}
 & $0.1\%$ & 0.163 & 0.019 & 4.425 \\
 & $1\%$   & 0.219 & 0.100 & 4.414 \\
 & $5\%$   & 0.275 & 0.269 & 4.414 \\
\midrule
\multirow{2}{*}{\emph{Exact} proprietary model}
 & $0.1\%$ & 0.250 & 0.275 & 4.430 \\
 & $1\%$   & 0.287 & 0.412 & 4.433 \\
\bottomrule
\end{tabular}
\end{table}

The results in Table~\ref{tab:lifecycle} confirm that reference quality matters. With a $0.1\%$ rollback budget, the three approximate references leave membrane/output BER at or below $0.169/0.019$. In contrast, the exact pre-embedding checkpoint increases BER to $0.250/0.275$ with almost no change in task loss. The reason is that the exact checkpoint isolates the changes made during watermark embedding much more closely. With the public initializer or an independently trained model, many of the largest parameter differences come from normal training rather than from MeMark, so the attacker is more likely to restore unrelated parameters. Thus, access to the exact pre-watermark checkpoint substantially weakens MeMark while leaving the task loss almost unchanged. This is a stronger attacker than the main threat model and is why this checkpoint has to remain private.

It is also worth noting that our public-reference experiment is an easier scenario for the attacker, since we obtain the proprietary model by adapting the public initializer for only 400 steps before embedding the watermark. A proprietary checkpoint that has undergone substantially more private training would likely contain more changes unrelated to MeMark, making parameter differencing against the public initializer less precise. We therefore view the robustness measured with the public initializer as a conservative estimate for this attack.

\section{Discussion and Limitations}
\label{sec:limitations}

MeMark is designed to identify checkpoint lineage, where the protected model is reused or modified. A fresh model obtained only from the outputs of the protected model is outside this setting. As shown in Appendix~\ref{app:extraction}, model extraction can reproduce the original task without preserving the watermark.

The rollback results also show that the exact checkpoint immediately before watermarking should remain private. If this model is available, the attacker can isolate the changes introduced during embedding much more accurately than with a public initializer or another related checkpoint.

The primary membrane watermark also requires white-box verification, since the verifier has to inspect the selected LIF states. The optional output response can be checked with only query access, but it can be removed by replacing the output head. Our threat model also assumes that the secret key and the exact pre-watermark checkpoint remain private. If either is fully disclosed, the attacker has substantially more information for targeting the watermark.

\section{Related Work}
\label{sec:relatedwork}

\textbf{SNN watermarking.} Previous SNN watermarking methods mainly verify ownership through the model output. Poursiami et al.~\cite{poursiami} adapt fingerprint and backdoor watermarks to SNNs, while SpikeTimer~\cite{spiketimer} uses temporal regions of the input for model authorization. 
However, their ownership evidence is tied to the model output. MeMark instead focuses on the internal spiking computation, so that the main watermark remains in the backbone when the output head is replaced.

\textbf{White-box watermarking.} White-box watermarks store ownership information in model weights or internal activations. Uchida et al.~\cite{uchida} embedded a payload directly in the weights, while passport methods use weights or normalization parameters~\cite{passport,passportnorm}. DeepSigns embeds information in internal activation distributions, and DICTION uses internal activations together with a learned projection to recover the payload~\cite{deepsigns,diction}. Other methods have similarly used hidden states or gates in recurrent networks~\cite{lim,gatekeeper}. 
MeMark also uses an internal state but leverages a property already present in SNNs. The payload is stored in selected membrane states, and each bit is read relative to the LIF neuron's own firing threshold. As such, our verification does not require a separate learned projection or reference value.

\textbf{Watermark removal and structural changes.} Previous work has studied watermark removal through model extraction and model modification~\cite{entangled,deepeclipse}, as well as structural transformations that reorganize the model while preserving its behavior~\cite{structuralobfuscation}. Other works show that corresponding neurons can be recovered by aligning their behavior~\cite{neuronalignment,findthelady}. These attacks are related to our structural experiments in Section~\ref{sec:permutation}, where changing the channel order moves the watermark coordinates without necessarily removing the neurons themselves.

\textbf{False ownership claims.}
Watermarking must also consider a malicious claimant. Fan et al.~\cite{passport} study ambiguity attacks, where an attacker forges a counterfeit ownership credential to cast doubt on the legitimate owner's claim. More recently, Liu et al.~\cite{mor} showed that model ownership verification can also be attacked by malicious accusers that construct ownership evidence for models they do not own. MeMark faces a similar problem: with white-box access, a claimant can inspect an unwatermarked model and choose coordinates and bits that already match its internal states. We therefore require the key to be committed before the protected checkpoint is released. This allows the verifier to distinguish a key registered by the owner before release from one selected afterwards.
\section{Conclusions}
\label{sec:conclusion}

In this work, we present MeMark, a watermark designed around the internal dynamics of SNNs. Instead of adding a separate decoder or storing the watermark in the output layer, MeMark uses the membrane potential of LIF neurons as its carrier. During training, the continuous distance to the firing threshold provides the signal needed to correctly embed the payload. After training, the watermark is recovered from the neuron's threshold decision, allowing the same mechanism that produces spikes in the original network to also provide watermark verification. This also makes verification lightweight, as MeMark does not require an additional learned projection to recover the payload, but only the selected membrane states and firing thresholds already present in the model.

Our experiments show that this provides a practical ownership signal for reused SNN checkpoints. The main advantage of MeMark is that the ownership response is read from the spiking backbone itself; it survives output-head replacement, does not require an internal learned projection, and reveals less of the registered payload than the matched activation projection when the challenge is not present.
On the 215.4M-parameter SpikeGPT model, all 20 independently embedded keys satisfy the fixed 51/64 verification rule, while none of the $30\,000$ fresh random keys pass it. All 20 keys also remain verifiable after 200 fine-tuning steps, $90\%$ pruning, int8 quantization, and replacement of the output head.

Structural changes can move the secret coordinates without removing the watermark itself, and we show that these coordinates can be recovered by aligning the suspect checkpoint with the protected release. We also find that attacks which search for the carrier neurons or modify membrane states without knowing the key can weaken the response, but do not remove the ownership evidence in the settings we test. In contrast, the rollback experiment shows that access to the exact pre-watermark checkpoint gives the attacker a much stronger removal reference. The same membrane-space mechanism can also be embedded across recurrent, convolutional, residual, and transformer SNNs, although the embedding settings need to be adjusted for the architecture.

We also show that the watermark response alone cannot establish ownership, as a claimant with white-box access can choose a key that already matches the existing model. Following prior ownership-resolution work, we require the key and protected release to be committed before publication. 

Future research could expand the proposed method and study whether the ownership signal can be preserved when a new model is obtained through distillation or model extraction, where the original internal states are no longer directly reused. It could also explore whether the embedding timeline and prior key registration can be incorporated directly into the model watermark, instead of relying on external mechanisms.

\bibliographystyle{plain}
\bibliography{Bibliography}

\appendix

%

\section{Watermark Key Generation}
\label{app:keygen}

The secret challenge is generated independently for each watermark key, and the same challenge is used to read all of its selected coordinates. For the recurrent and SpikeGPT models, $x_{\mathrm{wm}}$ is generated by sampling token identifiers uniformly from the model vocabulary, using 32 tokens for the recurrent model and 64 for SpikeGPT. For the vision models, we generate one fixed random input tensor whose values are sampled independently and uniformly from $[-1,1]$.

For all architectures, we first define the set of LIF states available to the watermark and then sample $B$ distinct pre-fire states uniformly without replacement. Thus, each payload bit is assigned to a different carrier state and every eligible state has the same probability of being selected.

For the recurrent model, all pre-fire states in the four recurrent LIF layers are eligible. For SpikeGPT, we use all 36 LIF modules. In the vision models, we use the spiking stages of the convolutional networks, the \texttt{sn1} and \texttt{sn2} nodes in \texttt{layer1} of SEW-ResNet-18, and the \texttt{lif1} and \texttt{lif2} nodes in each transformer block. All channels and timesteps within these modules are included in the eligible set.

\section{Additional Experimental Results}
\label{app:fullnumbers}

\subsection{Evaluation Testbeds}
\label{app:testbeds}
Table~\ref{tab:testbeds} summarizes the model architectures and watermark settings used in our experiments.

\begin{table*}[t]
\centering
\footnotesize
\setlength{\tabcolsep}{3.5pt}
\caption{Evaluation testbeds and watermark settings.}
\label{tab:testbeds}
\begin{tabular}{@{}llllrr@{}}
\toprule
Dataset or task & Architecture & LIF timesteps or context & Parameters & Embed steps & Payload \\
\midrule
enwik8 language modeling & 4-layer recurrent SNN & context 128 & 0.97M & 250 & 32 \\
OpenWebText checkpoint, enwik8 embed & SpikeGPT-216M & context 1\,024 & 215.4M & 2,000 & 64 \\
MNIST & 3-stage spiking CNN & 4 & 0.21M & 200 & 24 \\
CIFAR-10 & SEW-ResNet-18 & 4 & 11.18M & 300 & 16 \\
CIFAR-10 & Spikformer & 4 & 0.65M & 3\,000 & 16 \\
CIFAR-10 & QKFormer & 4 & 0.14M & 3\,000 & 16 \\
CIFAR10-DVS & 4-stage spiking CNN & 8 & 0.28M & 300 & 16 \\
N-Caltech101 & 4-stage spiking CNN & 16 & 1.13M & 800 & 16 \\
\bottomrule
\end{tabular}
\end{table*}

\subsection{Verification Threshold}
\label{app:fixed400}




We use the first five SpikeGPT keys to select both the embedding budget and the verification threshold. The keys are evaluated at $1\,200$, $2\,000$, and $3\,000$ steps using the same negative calibration population. Table~\ref{tab:budgetcal} reports the weakest genuine key and strongest negative key at each point.

\begin{table}[t]
\centering
\footnotesize
\caption{Calibration results for the first five SpikeGPT keys.}
\label{tab:budgetcal}
\begin{tabular}{@{}cccc@{}}
\toprule
Steps & Genuine min. & Negative max. & Difference \\
\midrule
 $1\,200$ &  53 &  49 &  4\\
 $2\,000$ &  59 &  48 &  11 \\
 $3\,000$ &  59 &  50 &  9\\
\bottomrule
\end{tabular}
\end{table}

The weakest genuine response improves from $1\,200$ to $2\,000$ steps and then remains unchanged at $3\,000$, so we use $2\,000$ steps for the main experiments. The strongest negative observed over the complete calibration reaches 50/64 bits. We therefore set the fixed verification threshold one bit above it, at 51/64.

\subsection{Embedding Steps}
\label{app:budget}

Here we compare the results of different step budgets. We consider 3 cases: first, the owner embeds the keys for 400 steps and keeps training the keys that failed on the fixed rule to $1\,200$ steps. The second setting is where the owner embeds all keys to $1\,200$ steps. And finally, we compare it to the $2\,000$-step budget. Table~\ref{tab:budgetfull} reports the complete 20-key results for the three checkpoint sets.

\begin{table}[ht]
\centering
\footnotesize
\caption{Effect of the SpikeGPT embedding budget. ``FT pass'' reports the number of keys that still pass after 200 fine-tuning steps.}
\label{tab:budgetfull}
\begin{tabular}{@{}lccc@{}}
\toprule
Training & Genuine range & Fresh FP & FT pass \\
\midrule
 400/$1\,200$ &  48--57 &  8/30\,000 &  19/20 \\
 all at $1\,200$ &  53--60 &  0/30\,000 &  19/20 \\
 all at $2\,000$ &  55--62 &  0/30\,000 &  20/20 \\
\bottomrule
\end{tabular}
\end{table}

The main difference appears after fine-tuning. At $1\,200$ steps, one key drops from 53 to 45 bits and no longer passes. At $2\,000$ steps, all 20 keys remain above the 51/64 rule, with the weakest response at 53 bits. The average change is similar, so the longer embedding mainly removes the large drop observed for one of the shorter-trained keys.

\subsection{Additional Watermark Ablations}
\label{sec:ablations}
\label{app:ablations}

We first check whether embedding the watermark changes the normal spiking behavior of the recurrent network. The firing rate is $0.2005$ for the equal-budget control and $0.1966$ after watermarking. Thus, moving the selected membrane states across their thresholds does not produce a general increase in spike activity.

We next study how the watermark coordinates should be selected. Membrane states from nearby timesteps of the same neuron are related, so assigning them conflicting bits can make the watermark harder to embed. We repeat this experiment over eight neurons and three timestep differences in each of the four recurrent layers, giving 24 configurations. When the two selected states have the same target bit, the mean watermark loss is $0.0033$. When their target bits are opposite, it increases to $0.0186$, and the conflicting case has the larger loss in 23 of the 24 configurations. This shows that nearby coordinates from the same neuron are not completely independent and should be selected with care.

We then increase the payload size to test whether this becomes a practical limitation as more coordinates are used. Table~\ref{tab:capacity_appendix} reports the complete recovery results. The membrane watermark recovers every tested payload from 8 to 128 bits. The optional output response is exact up to 64 bits and recovers 127 of 128 bits for the largest payload.

\begin{table}[t]
\centering
\footnotesize
\caption{Payload-size ablation on the recurrent SNN. Entries report correctly recovered bits.}
\label{tab:capacity_appendix}
\begin{tabular}{@{}lcc@{}}
\toprule
Payload & Membrane & Output \\
\midrule
8   & 8/8     & 8/8 \\
16  & 16/16   & 16/16 \\
32  & 32/32   & 32/32 \\
64  & 64/64   & 64/64 \\
128 & 128/128 & 127/128 \\
\bottomrule
\end{tabular}
\end{table}

Finally, we test whether the payload needs to cross the real firing threshold. We keep both target values below the firing threshold and introduce a separate internal boundary between them. The membrane watermark still reaches BER $0.000$, and the selected neurons do not need to fire to store the payload. Therefore, the continuous membrane state can carry the information on its own. However, this variant requires an additional decoding boundary. MeMark instead uses the firing threshold already present in the neuron, which avoids storing or calibrating another reference value.

\subsection{LIF Parameter Sensitivity}
\label{app:perturbcost}

\begin{figure*}[t]
\centering
\begin{subfigure}{0.49\textwidth}
\centering
\includegraphics[width=0.9\linewidth]{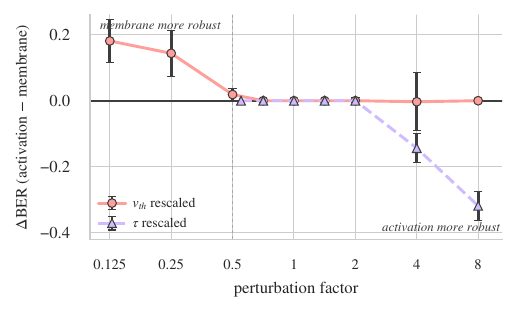}
\caption{MeMark and the matched DICTION-style activation watermark.}
\end{subfigure}\hfill
\begin{subfigure}{0.49\textwidth}
\centering
\includegraphics[width=0.9\linewidth]{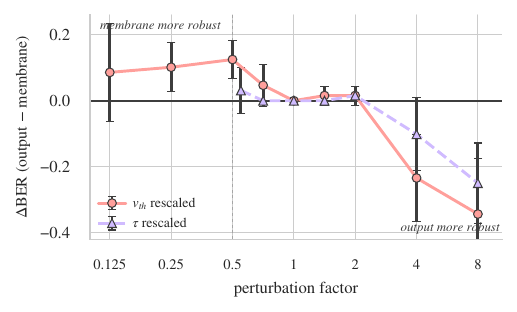}
\caption{Membrane and optional output responses.}
\end{subfigure}
\caption{LIF parameter changes. The methods react differently to large
changes in $v_{th}$ and $\tau$.}
\label{fig:lif_appendix}
\end{figure*}

Section~\ref{sec:carriermatrix} summarizes the effect of changing the LIF firing threshold and membrane time constant. Figure~\ref{fig:lif_appendix} shows how the relative robustness of the carriers changes across the sweep, while Table~\ref{tab:lif_full} reports the complete BER and validation-loss values. The MeMark and DICTION-style models start from the same checkpoint and are evaluated on the same validation batches. The models are not retrained during this experiment.

\begin{table}[t]
\centering
\footnotesize
\setlength{\tabcolsep}{2.6pt}
\caption{BER and validation loss under LIF parameter changes, mean over ten matched keys. $\Delta$ loss is measured from the unmodified model.}
\label{tab:lif_full}
\begin{tabular}{@{}lcccc@{}}
\toprule
Scale & MeMark & DICTION & $\Delta$ loss M & $\Delta$ loss D \\
\midrule
\multicolumn{5}{@{}l}{\emph{Firing threshold} $v_{th}$} \\
$0.125\times$ & 0.075 & 0.256 & +0.603 & +0.606 \\
$0.25\times$  & 0.034 & 0.178 & +0.510 & +0.513 \\
$0.5\times$   & 0.003 & 0.022 & +0.256 & +0.258 \\
$0.707\times$ & 0.000 & 0.000 & +0.088 & +0.088 \\
$1\times$     & 0.000 & 0.000 & +0.000 & +0.000 \\
$1.414\times$ & 0.000 & 0.000 & +0.087 & +0.085 \\
$2\times$     & 0.003 & 0.003 & +0.367 & +0.368 \\
$4\times$     & 0.366 & 0.362 & +1.356 & +1.366 \\
$8\times$     & 0.512 & 0.512 & +1.785 & +1.789 \\
\midrule
\multicolumn{5}{@{}l}{\emph{Time constant} $\tau$} \\
$0.55\times$  & 0.000 & 0.000 & +0.138 & +0.137 \\
$0.707\times$ & 0.000 & 0.000 & +0.050 & +0.050 \\
$1\times$     & 0.000 & 0.000 & +0.000 & +0.000 \\
$1.414\times$ & 0.000 & 0.000 & +0.049 & +0.047 \\
$2\times$     & 0.000 & 0.000 & +0.186 & +0.187 \\
$4\times$     & 0.153 & 0.009 & +0.546 & +0.547 \\
$8\times$     & 0.419 & 0.100 & +0.970 & +0.974 \\
\bottomrule
\end{tabular}
\end{table}

Lowering the firing threshold affects the activation watermark more, while increasing the membrane time constant affects MeMark more. However, the larger changes also have a substantial cost on the original task. For example, increasing $\tau$ to $8\times$ raises the membrane BER to $0.419$, but it also adds $0.970$ validation loss. Similarly, scaling $v_{th}$ by $8\times$ moves both watermarks close to chance and adds about $1.8$ loss. The watermark therefore starts degrading only after the LIF dynamics have already changed substantially.

\subsection{Partial Key Exposure}
\label{app:partialknowledge}

We explore the scenario where the challenge and part of the coordinate-bit pairs are revealed to the attacker. For every disclosed coordinate, the attacker knows both its location and its target bit and trains that state toward the opposite value for 250 steps. Table~\ref{tab:partial_appendix} separates the disclosed and undisclosed coordinates.

\begin{table}[t]
\centering
\footnotesize
\caption{Partial-key attack on the recurrent SNN.}
\label{tab:partial_appendix}
\begin{tabular}{@{}lcc@{}}
\toprule
Exposed fraction & Exposed BER & Hidden BER \\
\midrule
0.00 & --    & 0.000 \\
0.25 & 1.000 & 0.000 \\
0.50 & 1.000 & 0.000 \\
0.75 & 1.000 & 0.000 \\
1.00 & 1.000 & --    \\
\bottomrule
\end{tabular}
\end{table}

The attacker can flip every disclosed coordinate. However, the undisclosed coordinates remain at BER $0.000$ for all tested exposure rates. The separately keyed output response also remains at or near zero BER. This experiment tests whether removing the disclosed coordinates also damages the undisclosed ones.

\subsection{Watermark-Aware Event Timing}
\label{app:eventretiming}

The timing experiment in the main paper changes the event timestamps without using any information from the watermark. We also test a stronger attacker that can observe the watermark score while searching over the same temporal range.

With a budget of $50\,000$ queries, the membrane BER remains unchanged. Increasing the budget to $200\,000$ queries raises the membrane BER to $0.250$, while the output response remains at BER $0.000$. Thus, using the watermark score makes the timing search more damaging, but it still does not remove the registered response in the settings tested.

\subsection{Adaptive Structural Recovery}
\label{app:structuralfull}

The structural experiments in the main paper align channels using
behavioral responses on non-secret inputs. We also test whether an
attacker can make these responses harder to match. Here,
the attacker fine-tunes the suspect model with an additional objective that reduces the differences between the channel signatures while keeping the normal task objective.

\begin{table}[t]
\centering
\footnotesize
\caption{Recovery after modifying the channel signatures used for
alignment.}
\label{tab:adaptive_alignment}
\begin{tabular}{@{}lccc@{}}
\toprule
Condition & Val. loss $\Delta$ & Match acc. & Recovered BER \\
\midrule
Task $+$ signature flattening & $-0.008$ & 0.984 & 0.000 \\
Plain fine-tuning             & $-0.385$ & 0.828 & 0.031 \\
\bottomrule
\end{tabular}
\end{table}

Table~\ref{tab:adaptive_alignment} shows that the alignment remains accurate even when the attacker explicitly modifies the responses used for matching. Here, the attacker tries to make the channel responses more similar, so that it becomes harder to match them with the protected checkpoint. However, the alignment still matches $98.4\%$ of the channels correctly, and the recovered watermark remains at BER $0.000$. For comparison, after normal fine-tuning, the matching accuracy drops to $82.8\%$, while the recovered watermark has BER $0.031$. Thus, even when the channel responses are modified, the alignment can still recover the watermark coordinates.

\subsection{Watermark Detection}
\label{app:detection}

\begin{figure}[t]
\centering
\includegraphics[width=\columnwidth,trim=0pt 8pt 0pt 10pt, clip]{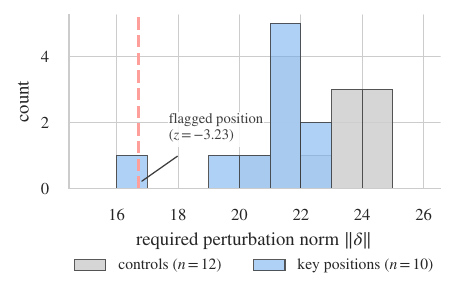}
\caption{Minimum-norm reconstruction for watermark detection. The
keyed positions largely overlap with the random controls, and only one
keyed position is flagged.}
\label{fig:detection_appendix}
\end{figure}

We repeat the minimum-norm reconstruction attack used in Section~\ref{sec:adaptive}. The attack can optimize a candidate input for all 22 tested positions, so successful reconstruction alone does not identify a watermark coordinate. As shown in Figure~\ref{fig:detection_appendix}, the reconstruction norms of the keyed positions largely overlap with the twelve random controls, and only one of the ten keyed positions is flagged.

We repeat the detector with five different seeds. The same flagged position is recovered in four out of five runs, while three representative keyed positions that were not initially flagged remain unflagged in every run. Thus, in this experiment the detector can find one unusual position, but it does not reliably separate the sparse watermark coordinates from normal membrane states.

\subsection{A Second Watermark}
\label{app:ambiguity}

Finally, we test whether adding another watermark removes the one that is already present. We embed an independently generated second key into a protected model and then verify the original key again.

On the recurrent SNN, the original membrane and output responses remain at BER $0.000/0.000$. We observe the same behavior on the
215.4M-parameter model, where the original responses remain at their pre-embedding values while the second key is added. Therefore, two watermark responses can coexist in the same checkpoint.

This experiment shows that adding a second watermark does not automatically erase the first one. It does not allow the model response itself to decide which claimant has priority, since both keys may verify on the same checkpoint. The ordering must therefore come from the commitments created before the corresponding claims. In other words, the watermark provides the model-level response, while the pre-release commitment provides the evidence of when that key was registered.

\subsection{Output Watermark}
\label{app:decoding}

The optional output watermark depends on the information available from the model output. When the output scores are available, each bit is associated with two possible outputs and is recovered by comparing their scores. The bit is read based on which of the two outputs has the higher score.

For interfaces that only return a predicted label or generated token, we instead use the selected labels or tokens directly. In this case, the response is correct when the model produces the output associated with the expected bit. This makes the output watermark possible to verify without reading the internal LIF states, but also makes it more dependent on the output interface. The membrane and output payloads are decoded separately and do not need to contain the same bits.

\subsection{Training Settings}
\label{app:pareto}

For the recurrent SNN and SpikeGPT experiments, we use $\lambda_{wm}=1$, $\lambda_{fid}=1$, and $\tau_{wm}=0.5$ as the default setting and keep these values fixed across the different keys rather than tuning them individually. The transformer models are more sensitive to the fidelity objective, so we only reduce $\lambda_{fid}$ for these architectures. For Spikformer, we use $\lambda_{fid}=0.25$, while for QKFormer we remove the fidelity term and use $\lambda_{fid}=0$. Both transformer models are trained for $3\,000$ embedding steps, as reported in Table~\ref{tab:testbeds}.

The remaining model-specific training settings, including the optimizer, learning rate, batch size, random seeds, and complete configurations, are provided in our repository.

\subsection{Model Extraction as a Claim Boundary}
\label{sec:entangled}
\label{app:extraction}

The attacks in our threat model start from the protected checkpoint and modify or reuse its internal computation. Model extraction is different because the attacker can instead use the protected model only as a teacher and train a fresh student. We include this experiment as a boundary of the ownership claim rather than as another robustness attack. The student is trained on normal inputs and never observes the secret challenge or the protected membrane states. After $1\,500$ distillation steps, its task loss reaches $2.447$, compared with $2.442$ for the teacher, showing that the public task behavior has been copied.

However, the watermark does not transfer with this behavior. The student reaches membrane BER $0.531$ and output BER $0.438$, both close to random. This is expected, as the distillation objective only matches normal model outputs and does not contain any information about the secret internal response. Thus, MeMark can identify reuse and modification of the protected checkpoint, but not a fresh model obtained only through output-based extraction.

\subsection{Verification Cost}
\label{app:verificationcost}

Table~\ref{tab:deploy} reports the complete verification-cost measurements for the $215.4$M-parameter SpikeGPT checkpoint. The measurements use 256-token sequences and report the median of 30 runs on an RTX4090.

\begin{table}[t]
\centering
\footnotesize
\caption{Verification cost on the 215.4M-parameter SpikeGPT checkpoint,
measured with 256-token sequences and the median of 30 runs on an RTX 4090.}
\label{tab:deploy}
\begin{tabular}{@{}ll@{}}
\toprule
Quantity & Measured \\
\midrule
Checkpoint / secret key size & 861.7\,MB / $1\,600$\,bytes \\
Membrane state on the challenge & $1\,769\,472$ scalars \\
Scalars the verifier reads & 64 ($0.0036\%$) \\
LIF modules that must be observable & 30 of 36 \\
\quad with the keyed modules tapped & 788\,ms ($+1.9\%$) \\
\quad with all LIF modules tapped & 791\,ms ($+2.3\%$) \\
End-to-end, including checkpoint load & 2.59\,s \\
Strict black-box path & 64 queries \\
Firing rate vs pretrained & $+0.67\%$ \\
\bottomrule
\end{tabular}
\end{table}

\end{document}